\documentclass[aps,pra,twocolumn,groupedaddress, longbibliography]{revtex4-1}

\usepackage{graphicx}
\usepackage{subfigure}
\usepackage{amsmath}
\begin{document}

\title{Iteratively seeded mode-locking}

\author{$^{1}$Victor G. Bucklew, $^{2}$William H. Renninger, $^{1}$Perry S. Edwards, $^{1,3}$Zhiwen Liu}

\affiliation{$^{1}$Atoptix Inc, 200 Innovation Blvd., State College PA 16803, USA\\
$^{2}$Department of Applied Physics, Yale University, New Haven, Connecticut, 06520, USA\\
$^{3}$Department of Electrical Engineering, The Pennsylvania State University, University Park, PA, USA}

\date{\today}

\begin{abstract}
Ultrashort pulsed mode-locked lasers enable research at new time-scales and revolutionary technologies from bioimaging to materials processing. In general, the performance of these lasers is determined by the degree to which the pulses of a particular resonator can be scaled in energy and pulse duration before destabilizing. To date, milestones have come from the application of more tolerant pulse solutions, drawing on nonlinear concepts like soliton formation and self-similarity. Despite these advances, lasers have not reached the predicted performance limits anticipated by these new solutions. In this letter, towards resolving this discrepancy, we demonstrate that the route by which the laser arrives at the solution presents a limit to performance which, moreover, is reached before the solution itself becomes unstable. In contrast to known self-starting limitations stemming from suboptimal saturable absorption, we show that this limit persists even with an ideal saturable absorber.  Furthermore, we demonstrate that this limit can be completely surmounted with an iteratively seeded technique for mode-locking.  Iteratively seeded mode-locking is numerically explored and compared to traditional static seeding, initially achieving a five-fold increase in energy. This approach is broadly applicable to mode-locked lasers and can be readily implemented into existing experimental architectures.
\end{abstract}

\maketitle

Mode-locked laser systems generating ultrashort pulses with exceptional performance qualities (e.g. high-energy,  short temporal duration, high peak powers) are attractive for countermeasure applications, nonlinear imaging, materials characterization and processing, and fundamental studies involving frequency comb metrology and the understanding of ultrafast dynamics \cite{richardson2010}.

Achieving exceptional performance qualities presents a significant challenge because the nonlinear dynamics which underlie pulse formation in a laser resonator are complex.  Major advances for ultrashort-pulsed laser development have come through new understandings of the steady-state behavior of pulse evolutions.  This is particularly evident in fiber laser systems with the development of stretched-pulse \cite{tamura1993}, passive self similar \cite{ilday2004self, bucklew2012realizing}, amplifier similariton \cite{renninger2010self, oktem2010soliton, agaray2010experimental}, and dissipative soliton evolutions \cite{soto1997pulse, grelu2012dissipative, renninger2008, chong2007, chong2006, kalashnikov2006, ding2012high,buckley2007stabilization, bale2008spectral, wise2008high, cabasse2009high}, demonstrating that pulse qualities can be altered or optimized through careful engineering of resonator characteristics.

Recent developments in algorithmic approaches to mode-locking have helped to further propel the field by optimizing the multi-parameter design space of these resonators in a way that is difficult or impossible through manual design.  Specifically, researchers have implemented algorithmic and machine learning approaches to resonator parameter control, which, in concert with a suitable figure of merit, can help optimize a resonator for a certain pulse quality such as a minimum pulse width or a high peak power \cite{woodward2016self, haefner2016rigorous,  woodward2016towards, andral2016toward, andral2015fiber, brunton2014self, fu2013high, brunton2013extremum, ilday2016}.

Despite these major advances, evidence suggests that experiments have not yet succeeded in achieving the highest performance qualities anticipated theoretically \cite{haus1995stretched, mollenauer1984soliton, haus1991structures, renninger2012pulse, bucklew2014average, renninger2010area}.  For example, the phenomenon referred to as a dissipative soliton resonance predicts near limit-less enhancement of the pulse energy in a dissipative soliton laser, given the appropriate laser design \cite{akhmediev2008, wu2009dissipative}.  However, although experimental observations of these resonances have been published, the discrepancy between theory and experiment suggests that we have just begun to tap into the full potential which these pulse types and others can theoretically offer.

In this letter, to help resolve the disparity between steady-state mode-locking theory and experimental implementation, we demonstrate a new limit to mode-locking based on the route by which the pulse arrives the steady-state solution.  Furthermore, we demonstrate that this limit can be completely surmounted with a new method of mode-locking based on iteratively seeding the resonator, which we refer to as iteratively seeded mode-locking (ISM).  In a particular example, a five-fold increase in energy is demonstrated.

To date, mode-locked laser systems have been largely understood by concentrating on the required relationship between a desired set of pulse characteristics and the final resonator state that the pulse exists within, or the steady-state solution.  More generally, however, the generation of ultrashort pulses is a gradual transition from an initial state into a final steady state.   Moreover, a desired final pulse state can only be realized within a specific resonator if that initial pulse state lies within the basin of nonlinear attraction of the final pulse state.

Part of this route to the steady-state solution is well known to laser scientists.  The concept of 'self-starting', for example, in which a laser successfully mode-locks from a noisy initial condition has been very well studied.  Many laser systems, such as solid-state lasers, do not mode-lock directly from noise and require additional intervention, such as with acousto-optic modulators \cite{hargrove1964locking}, or even by directly tapping the table \cite{spence1991}.  In fiber lasers, for example, the wave-plates of a nonlinear polarization evolution laser must rotated until the laser successfully initiates.  In general, to date, the concept of self-starting refers to the quality of the saturable absorber (the 'mode-locker').  If the saturable absorber is insufficient to bring the optical field from noisy fluctuations to short enough fluctuation to be affected by dispersion and nonlinearity, the laser will not 'self-start'.  In this work, we examine a new limit which occurs even if the optical field was initialized with a pulse.  In other words, this new limit occurs even with a ideal saturable absorber.

Regions of nonlinear attraction (and the range of initial optical states that a region can stabilize into a mode-locked state) can vary substantially depending on the desired characteristics of the mode-locked state.  This is a consequence of the complicated nonlinear landscape that determines basins of attraction.  Once pulse formation is initiated with an appropriate saturable absorber, research examining how to suppress unwanted solutions (e.g. multi-pulsing instabilities \cite{ding2012high,li2010geometrical}) and select desired pulse solutions within a complicated resonator phase space of competing pulse solutions, demonstrates the importance of designing a resonator to support pulse formation within a desired basin of attraction.  Research examining how to improve pulse states through programmable wave plates, demonstrates the effect that relatively small one-dimensional changes to cavity elements can have in the local optimization of a pulse quality \cite{woodward2016self, haefner2016rigorous,  woodward2016towards, andral2016toward, andral2015fiber, brunton2014self, fu2013high, brunton2013extremum}.  Recent research demonstrating reversible and irreversible pulse state transitions through use of an SLM inside of a resonator \cite{ilday2016}, and observation of hysteresis effects \cite{liu2009multistability}, point to the importance which the path taken to a potential solution can have in stabilization of that solution.

We propose that by understanding these developments to mode-locking through the context of both the potential steady-state solution, and also the path taken to reach that solution, a new limit to high performance pulse generation can be identified.  We show that through this lens of understanding, precise control over both the roundtrip resonator parameter space, as well as the seed state for each point along the path taken to a high performance mode-locking solution, is required.  By accounting for both, it is possible to closely follow a desired region of nonlinear attraction (ideally corresponding to the exact solution of the resonator) while traversing a multi-dimensional resonator phase space, and thereby stabilize pulse states which are not accessible through statically seeded resonator designs.

Specifically, as can be seen in Fig 1b, by stepping through a multi-dimensional resonator parameter space (the temporal trajectory of which is denoted by the black line), it is possible to dynamically and closely follow the exact steady state solution of the resonator at each new step in the resonator phase space (shown by the dotted white line).  In so doing, the steady state pulse supported by each previous resonator step can be used as an effective seed state for the new resonator configuration (as it also lies within a window of nonlinear attraction for the new resonator state).  Fig. 1c shows the pulse state as a function of round trip number for a conventional static-seeded approach which uses a single optical state to seed a static resonator, in order to reach a desired steady state pulse solution with properties very different from the initial seed state.  For reaching high performance mode-locking states far from the initial state of a laser, a static seeded approach can not stabilize pulse formation.  In contrast, Fig 1d shows that by iteratively changing the resonator state of a laser in the way just described, the optical state produced in the previous step and which thereafter seeds the current resonator step, is very close in quality to the pulse state of the new resonator step, allowing the successful formation and stabilization of that new pulse state.  By following the ISM method, we suggest that it is possible to practically realize theoretically anticipated high performance mode-locking states not accessible through conventional techniques which can not, by virtue of their static designs, provide the required resonator conditions needed to stabilize these more extreme states.

\begin{figure}[ht]
     	\centering
           \includegraphics[width=1\linewidth]{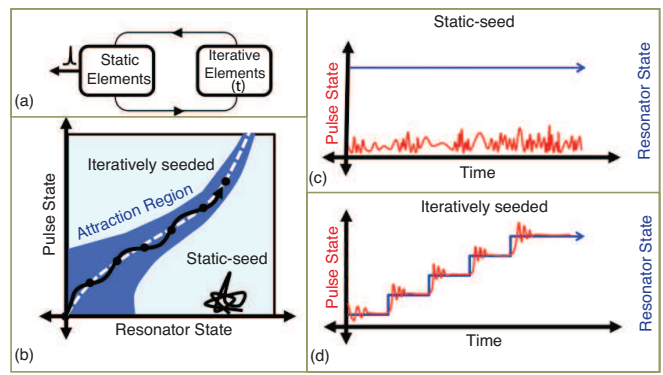}\\
           \caption{[Color online] (a) The elements in the laser cavity are iteratively changed as a function of time in order to stabilize an evolving pulse. (b) Reaching a pulse state requires that the initial state seeding the pulse evolution lies within a region of attraction (dark blue) of the exact solution (white line) supported by that resonator state.  Regions of attraction denote areas where a seed state can be pulled in to the steady state solution that is linked to that specific resonator state (x-axis).  By incrementally changing the resonator state, the pulse generated in the previous resonator state can be made to lie within a region of attraction for the new state, and thus safely be transitioned into a new steady state pulse solution (y-axis).  In standard designs when statically seeded states do not lie within an attraction region of the desired final resonator state, pulse formation is not observed. (c) The pulse state of a standard statically seeded mode-locked laser is shown as a function of time.  (d) The pulse state and resonator state of an iteratively-seeded mode-locked laser with the same final resonator state as (c) are shown as a function of time.}
\label{vision}
\end{figure}

To explore the hypothesis in more detail, the ISM method is numerically applied to a dissipative soliton laser for the optimization of pulse energy.  We use a standard split step Fourier method to solve a generalized Nonlinear Schrodinger Equation which has been validated as a useful method and model for describing dissipative soliton pulse evolution in mode-locked lasers \cite{agrawal2007nonlinear, renninger2012pulse,  grelu2012dissipative}.  Full details on the numerical simulation can be found in the Suppl. Mat 1.  A perfect saturable absorber is used in simulations in order to demonstrate that the proposed limit is not only a way to mitigate concerns of reaching a region of nonlinear attraction in the first place (stemming from imperfect saturable absorption), but represents a more fundamental limit that is operable even when pulse formation is fully supported by a perfect saturable absorber.

\begin{figure}[ht]
     	\centering
           \includegraphics[width=1\linewidth]{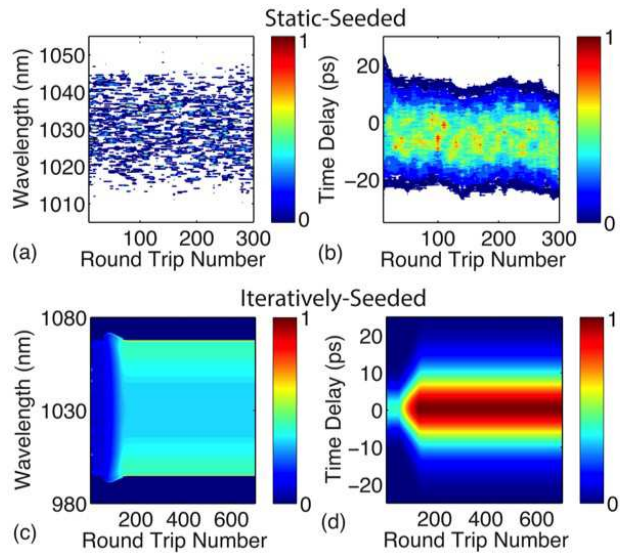}\\
            \caption{[Color online] Representative simulations of the spectral and temporal evolutions of a pulse for a resonator with the same final cavity parameters. (a) and (b) represent a standard static seeded resonator. (c) and (d) represent an iteratively seeded resonator.}
     \label{fig2}
      \end{figure}

Fig 2 shows results of representative simulations of a static seeded resonator and an ISM resonator.  The final cavity parameters of the static-seeded and ISM resonators are the same, although the parameters of the standard noised-seeded resonator do not change while the ISM parameters do.  For the ISM simulation, the pulse is initially allowed to build from noise in a static seed setting known to support pulse formation.  After the pulse has formed in this configuration (represented in Fig 3b and Fig 3d by the point of origin), the saturation energy of the resonator and the added GDD of the resonator are increased in a fixed linear relationship until a final pulse energy and resonator state is reached.  The slope of this relationship for a particular cavity can be seen by connecting a line from the origin of Fig 3d to the dot representing the final configuration of the resonator.  After this point, both the GDD and saturation energy are held fixed and the cavity becomes static again.  As seen in Fig 2c and Fig 2d an ISM resonator supports pulse formation whereas a static resonator does not (See Fig 2a and Fig 2b).  Stability is assessed by ensuring that the output pulse energy does not change for more than 200 round trips.

\begin{figure}[ht]
     	\centering
           \includegraphics[width=1\linewidth]{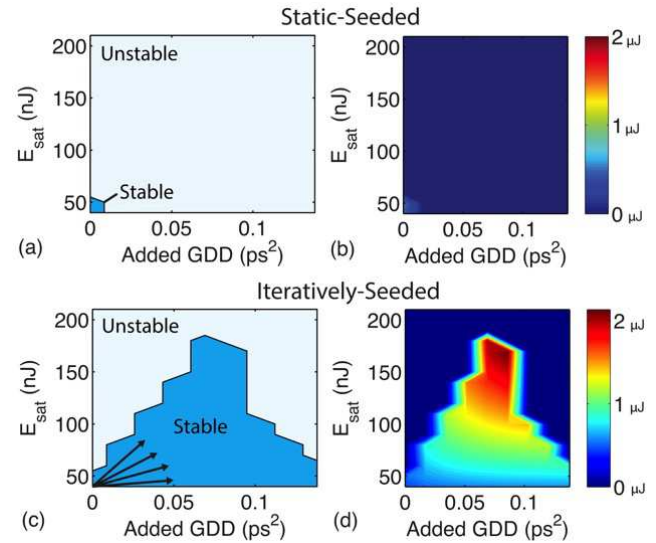}\\
           \caption{[Color online] Map of cavity configurations for a standard noise seeded (a-b) and an ISM (c-d) cavity.  Dark blue regions in (a) and (c) denote stable cavity configurations whereas light blue regions in (a) and (c) denote configurations that do not produce stable pulse evolutions.  The ISM simulations begin at the origin dot in a linear trajectory in (a) and (c) until reaching a designated end point (The figure represents results of 162 simulations arranged in a 9 x 18 point grid of GDD x E$_{sat}$ for each cavity type). Figures (b) and (d) represent energy contours of these simulations, showing that in this example, an ISM design can generate pulses with 5x more energy than in similar statically seeded designs initialized with either cavity noise, a broad several hundred picosecond long pulse representative of acousto-optic seeding, or picosecond scale cavity fluctuations reflective of table tapping.}
     \label{fig3}
\end{figure}

The regions of stability shown in Fig 3 show that there is a much larger range of cavity states where pulse formation is observed in an ISM resonator (Fig 3c) than in a static-seeded resonator (Fig 3a).  Correspondingly, the ISM resonator supports more extreme pulse states (e.g. higher energy, larger bandwidth.  For example, Fig 3d, shows a five time increase in pulse energy in this representative cavity system when mode-locked with an ISM design.

Now that a representative demonstration of the potential for ISM to approach extreme pulse states has been shown, we more closely examine the underlying principles for why ISM is successful.  To gain more insight into the pulse dynamics between and during resonator steps, numerical experiments are performed where after every change in the resonator parameters, the cavity is held static to see if the evolving pulse settles into a steady state for that specific cavity configuration.  An observation of settling into a steady state reflects that the pulse state of the previous resonator step lies within a region of of nonlinear attraction of the new resonator step. This means that stabilization is representative of the new resonator state pulling the previously stabilized pulse state, which is near to the exact solution of the resonator, into the new exact solution that it supports.

Fig 4a and Fig 4c show the results of the temporal evolution of optical radiation in a statically seeded and an ISM laser with the same final cavity configurations.  Fig 4d shows that the energy of the ISM resonator incrementally changes and stabilizes after each resonator step, whereas in Fig 4b, the energy never stabilizes.  To further clarify this observation, a pulse quality metric Q is presented.  The pulse quality Q is 1 if a well-defined single pulse peak is present in the cavity and asymptotically approaches 0 for pure noise.  The metric is defined as the inverse of the number of peaks with powers greater than 5$\%$ or more of the highest peak power in the cavity.  Numerical results of the pulse quality are computed for the noise-seeded and ISM cavity.  As anticipated, for the ISM resonator the pulse quality is 1 (See Fig 4d), representing a well-defined pulse, whereas for the standard static seeded resonator (See Fig 4b), the pulse quality is 0, showing that no pulse formation is observed.  The specific route taken to a final steady state does matter, especially as pulse characteristics become more extreme, and is further explored in the Supplement to this Letter.  As expected, the ISM simulations show that every resonator step is such that the previous pulse state, lying with a region of attraction, can be drawn into a stable steady state solution for that resonator state.  This suggests that if one could reproduce the theoretically predicted steady state cavity-pulse state relationship of the master equation explaining a specific pulse evolution through an iteratively seeded mode-locking design, it would be possible to generate pulses with sought after characteristics (e.g. high energy, short temporal widths, high peak power, quadratic spectral phase profiles).

\begin{figure}[h t]
	\centering
           \includegraphics[width=1\linewidth]{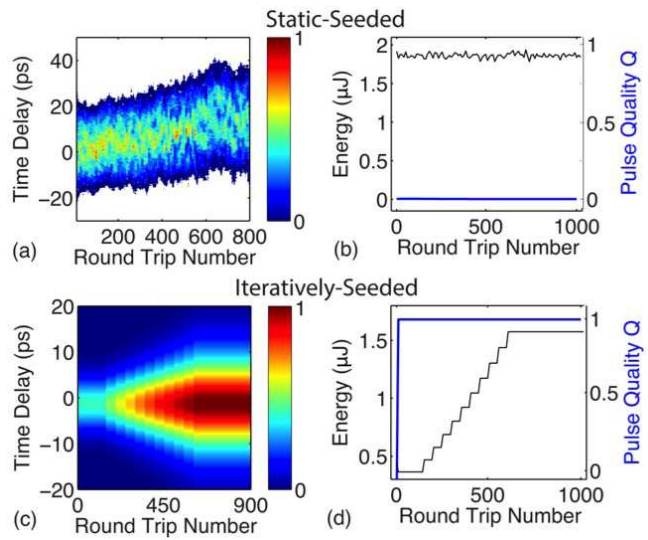}\\
               \caption{[Color online] After every change in cavity configuration for an evolving pulse in the ISM cavity, the pulse was allowed to settle before taking the next step.  These simulations show that each cavity step in the ISM resonator is able to stabilize and pull the pulse in the previous cavity step into a steady state solution.  A comparison with a static-seeded resonator is shown to demonstrate that at no point in the evolution of the static seeded system is a mode-locked state stabilized. (a) Temporal evolution of a pulse as a function of round trip in a static-seeded resonator; (b) Energy and pulse quality Q as a function of round trip number for a static-seeded resonator; (c) Temporal evolution of a pulse as a function of round trip in an iteratively-seeded resonator; (d) Energy and pulse quality Q as a function of round trip number for an iteratively-seeded resonator.}
     \label{fig4}
\end{figure}

Simulations show that the rate of change of resonator elements is not critical.  However, for very fast rates of change (on the order of a few round trips) between initial and final resonator states, a pulse does not have adequate time to adjust itself before the resonator state changes (and the corresponding window of nonlinear attraction), and will not form.

An ISM laser could be constructed with a feedback system and electronic control over cavity elements (Fig 1a) similar to those used in algorithmic mode-locking techniques. However, the emphasis of the iteratively seeded system is on both the desired final pulse state and the route taken to get there.  Stable parameter spaces will thus be identified through linked chains of cavity states rather than by single optimal states.  Comparing experimentally achievable electronic rates of change of cavity elements (i.e. on $\mu$s scale), with nanosecond laser round trip times, a pulse has plenty of time to stabilize between cavity states and the timing of resonator steps is not be crucial.  Methods already exist for tuning resonator components such as the energy, dispersion, saturable absorber, and spectral filter \cite{woodward2016self, haefner2016rigorous, woodward2016towards,andral2016toward, andral2015fiber, brunton2014self, fu2013high,brunton2013extremum}.  A recent demonstration of control over mode-locking states with a programmable spatial light modulator inside of a laser resonator \cite{ilday2016} demonstrates that robust experimental architectures exist and have been developed which are capable of implementing ISM designs.  It is anticipated that enabling an ISM approach with established algorithmic mode-locking techniques will help to extend the experimental operating regimes of ultrafast laser systems.

In conclusion, this letter identifies a new limit of high performance mode-locking by identifying a fundamental condition that must be met for pulse generation to be experimentally observed.  We show that the path taken to a desired final state is just as important as the final state itself, and that pulse formation will not be observed if the resonator environment cannot keep a pulse within a region of nonlinear attraction.  By incrementally changing the properties of a resonator so that after each step, the pulse produced by the previous resonator step lies within a region of nonlinear attraction of the new resonator state, it is shown that it is possible to stabilize pulse states which standard static-seeded laser designs do not admit.  Although the conclusions of this letter are directed towards broadening the scope of high performance mode-locking, they are not limited therein.  By identifying a new limit to the steady state performance of positive feedback systems, through consideration of both the desired final state, as well as the path required to reach that state, the conclusions of this letter provide a method (and understanding) that can be directly applied to expanding the operating regimes of all types of laser resonators.

Please see the corresponding supplementary material for more information.

\textbf{Funding:} Research reported in this publication was supported by the National Institute Of General Medical Sciences of the National Institutes of Health under Award Number R43GM113563. The content is solely the responsibility of the authors and does not necessarily represent the official views of the National Institutes of Health.

\providecommand{\noopsort}[1]{}\providecommand{\singleletter}[1]{#1}%

\end{document}


\title{Supplementary Material: Iteratively seeded mode-locking}

\author{$^{1}$Victor G. Bucklew, $^{2}$William H. Renninger, $^{1}$Perry S. Edwards, $^{1,3}$Zhiwen Liu}

\affiliation{$^{1}$Atoptix Inc, 200 Innovation Blvd., State College PA 16803, USA\\
$^{2}$Department of Applied Physics, Yale University, New Haven, Connecticut, 06520, USA\\
$^{3}$Department of Electrical Engineering, The Pennsylvania State University, University Park, PA, USA}

\date{\today}

\begin{abstract}
A description of the numerical methods presented in $\lq$Iteratively seeded mode-locking', is presented.  Parameters used in the simulations are identified and explained.  Additional results produced and enabled through the described numerical methodology are provided.\end{abstract}

\maketitle

Numerical simulations are based on modeling of a dissipative soliton laser resonator with a generalized nonlinear Schrodinger equation shown in Eq \ref{NLSE} within the gain medium in order to account for the interplay between the circulating optical pulse and the cavity elements within the resonator.  Although a laser designed to generate dissipative soliton pulses is used in the corresponding article, the presented approach of iteratively seeded mode-locking is applicable to other ultrafast pulse evolutions.

\begin{equation}
\begin{aligned}
\frac{\partial A(z,t)}{\partial z} = [\frac{g_{0}(z)}{1+\frac{E_{p}}{E_{sat}}}(1+\frac{1}{\omega_{c}^{2}}\frac{\partial^{2}}{\partial t^{2}}) - \frac{j}{2} \beta(z) \frac{\partial^{2}}{\partial t^{2}}\\
+ j \gamma(z) |A(z,t)|^{2}] A(z,t).\\
 \end{aligned}
\label{NLSE}
\end{equation}

\begin{figure}[h t]
               \centering
     \subfigure{
          \includegraphics[width=1\linewidth]{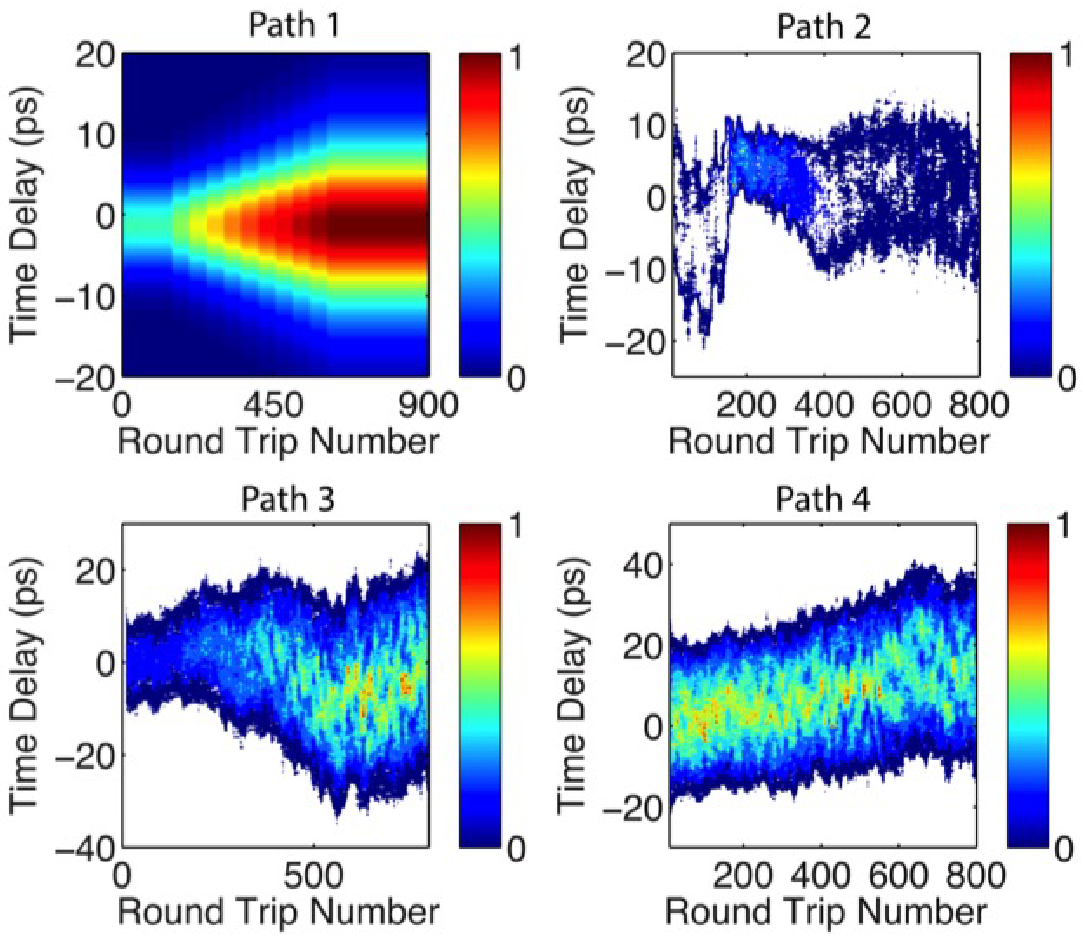}}\\
            \caption{[Color online] The route taken to a final resonator state determines whether a pulse will or will not form.  Here, the temporal evolution of a pulse is shown for four different cavity routes which each have the same final configuration but only one of which stabilizes pulse formation. \textbf{Path 1}: Cavity group delay dispersion is varied while saturation energy is varied; \textbf{Path 2}: Cavity group delay dispersion is varied while saturation energy is held fixed; \textbf{Path 3}: Cavity group delay dispersion is held fixed while saturation energy is varied. This case represents an analogue of gradually increasing the pump power in a noise-seeded laser cavity; \textbf{Path 4} Cavity group delay dispersion and saturation energy are both held fixed representing a standard noise-seeded cavity.}
            \label{S1}
\end{figure}

Eq. \ref{NLSE} is solved numerically with a standard split step Fourier algorithm \cite{agrawal2007nonlinear}. The simulation is seeded with a Gaussian noise sequence in the temporal domain and passed through a filter to smooth out femtosecond temporal fluctuations, allowing seed radiation to be more representative of real cavity noise. A(z,t) denotes the electric field envelope.  g$_{0}$(z) is the unsaturated gain, a piecewise constant function with value zero outside the gain medium.  E$_{p}$ is the intra-cavity pulse energy, which is updated after every round trip to account for gain saturation. $\omega_{c}$ corresponds to the full width at half maximum bandwidth (40 nm) of the Yb doped fiber gain medium. E$_{sat}$ is the saturation energy, normalized to the cavity round trip time, and is the mechanism in the simulation used to control the pulse energy. $\beta$(z), represents the group velocity dispersion parameter and is a piecewise constant function whose value is dependent on the medium the pulse is traveling within. Within the 3 m length of gain fiber $\beta_{g}$(z) is 23 fs$^{2}$/mm.  The additional dispersion which is added to the cavity either statically or dynamically as the pulse evolves is assumed through elements only providing dispersion and is thus represented in simulation through a lumped group delay dispersion (GDD) term.  $\gamma$(z) is also a piecewise constant function which is only non-zero within the gain medium and accounts for the accumulation of nonlinear phase.  Within the gain medium, $\gamma$ is 1.69E-3 m/W, and represents the adjusted nonlinear parameter for a 10$\mu$m large-mode-area double clad gain fiber used in the simulations.  

After exiting the gain medium, the circulating pulse encounters a lumped transmission loss of 80$\%$ to account for both the output coupling, as well as a lumped sum loss of other cavity elements.  Next, the pulse is passed through a saturable absorber modeled with a transmission function given as $T(|A(z,t)|^{2}) = 1-\frac{l_{0}}{1+|A(z,t)|^{2}/P_{sat}}$ \cite{renninger2012pulse} where $l_{0}$ is the unsaturated loss and $P_{sat}$ is the saturation power.  Here, $l_{0}$ is 1, which both helps demonstrate the proposed limit even in the absence of self-starting concerns resulting from imperfect saturable absorption, as well as provides stabilization against cavity noise fluctuations required for supporting high-energy pulse formation.  For example, it has been shown that large modulation depths support higher energy pulse states by adequate suppression of the continuous wave background on each round trip \cite{renninger2015fundamental}.  Although nonlinear polarization techniques can nearly approach this degree of modulation depth, further advances in increasing modulation depth of saturable absorbers are needed.  P$_{sat}$ is set to 8.4 kW, which represents an average saturation of the pulse peak power P$_{peak}$ to P$_{sat}$ of 3X (varying from 1X-10X) across all of the simulations.  After passing through the saturable absorber the pulse once again enters the gain medium for another cavity round trip.

For the iteratively-seeded simulations in the Letter, the pulse is initially allowed to build from noise for 60 round trips in a static seed setting (where the added GDD$_{added}=$0 and E$_{sat}=$40 nJ).  After the pulse has formed in this configuration, the saturation energy of the resonator and the added GDD of the resonator are increased in a fixed linear relationship for 80 cavity round trips.  The rate of change of the resonator parameters is not critical as long as the pulse has enough time to be adjust its properties to the new resonator configuration in between each cavity step.

\begin{figure}[h t]
               \centering
          \includegraphics[width=1\linewidth]{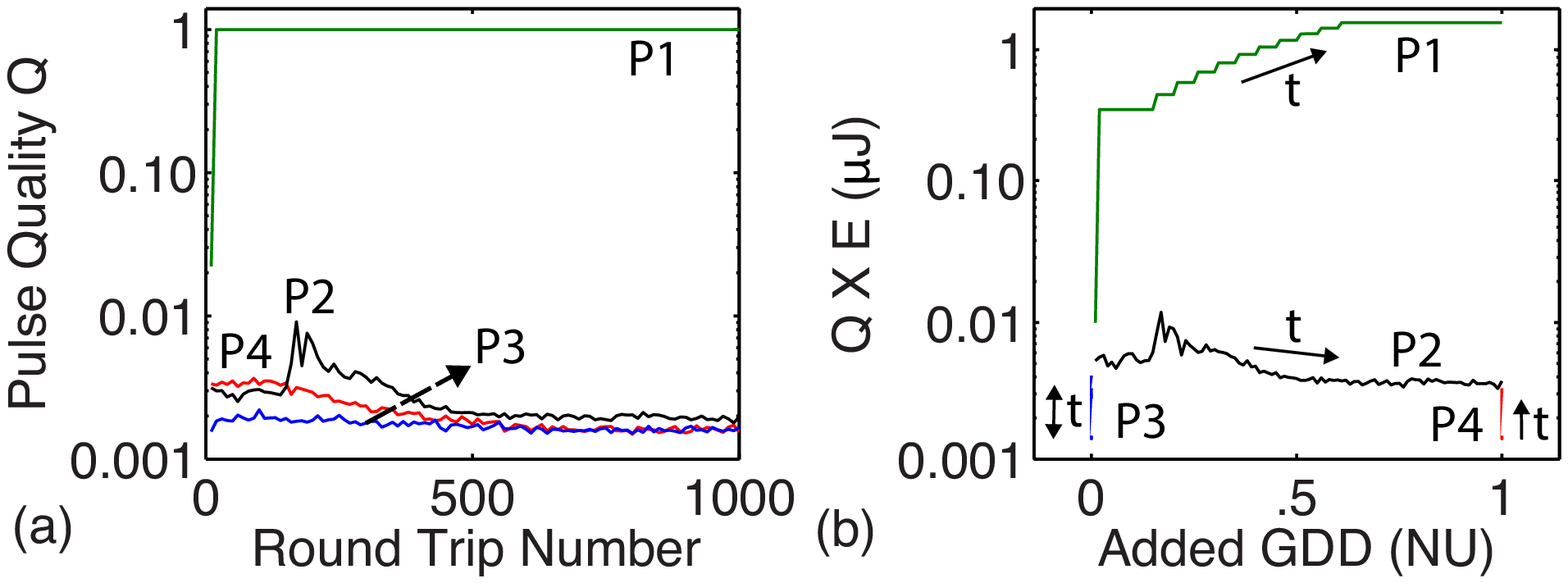}\\
            \caption{[Color online] (a) The pulse quality Q is shown as a function of round trip number for the four different cavity paths P1-P4 shown in Fig \ref{S1}.  (b) The pulse state (represented by the quantity Q X E) is shown as a function of cavity state (represented by the added group delay dispersion).  The arrows point in the direction of time.  Only Path 1, which varies both pulse energy and group delay dispersion at the same time, leads to extreme pulse state generation.}
            \label{S2}
\end{figure}

For one example which demonstrates that the specific cavity route taken to a final pulse state determines whether that pulse state will be reached or not, further numerical results are presented.  The temporal evolutions of four different cavity routes taken to a final resonator state are shown in Fig 1.  As can be seen, only Path 1, which linearly varies both saturation energy and cavity dispersion at the same time, allows for stable pulse formation.  The pulse quality parameter Q for Paths 1-4 are shown in Fig 2a as a function of round trip number which also shows that the Path 1 cavity route produces a well defined pulse at every cavity step, whereas the other cavity paths, which take different routes to the same final resonator configuration, do not lead to pulse formation.  For example, Fig 2b shows the product of pulse quality times pulse energy (Q X E$_{p}$) which serves as a metric of pulse state.  The metric Q X E$_{p}$ of pulse state is shown as a function of added group delay dispersion, which represents a metric for the cavity state.  Fig 2b shows that Path 1 is able to reach a sought after extreme pulse but Paths 2-4, which take different approaches to reach that state, never lead to pulse formation.  The simulations show that if a cavity route does not lie within a region of nonlinear attraction, pulse formation will not occur.

\textbf{Funding:} Research reported in this publication was supported by the National Institute Of General Medical Sciences of the National Institutes of Health under Award Number R43GM113563. The content is solely the responsibility of the authors and does not necessarily represent the official views of the National Institutes of Health.

\providecommand{\noopsort}[1]{}\providecommand{\singleletter}[1]{#1}%